\documentclass[twocolumn]{revtex4-1}
\usepackage{mathtools}
\usepackage{graphics}
\usepackage{amsmath}
\usepackage{graphicx}
\usepackage{hyperref}
\usepackage{epstopdf}
\usepackage{xcolor}
\usepackage[hypcap]{caption}
\begin{document}
\title{Design of a Compact Spectrometer for High-Flux MeV Gamma-Ray Beams}
\date{\today}

\author{D.J. Corvan}
\email{dcorvan01@qub.ac.uk}
\affiliation {School of Mathematics and Physics, The Queen'€™s University of Belfast, BT7 1NN Belfast, UK} 
\author{G. Sarri}
\affiliation {School of Mathematics and Physics, The Queen'€™s University of Belfast, BT7 1NN Belfast, UK} 
\author{M. Zepf}
\affiliation {School of Mathematics and Physics, The Queen'€™s University of Belfast, BT7 1NN Belfast, UK} 
\affiliation {Friedrich-Schiller-Universit\"{a}t Jena, F\"{o}stengraben 1, 07740 Jena, Germany}

\begin{abstract}
A novel design for a compact gamma-ray spectrometer is presented. The proposed system allows for spectroscopy of high-flux multi-MeV gamma-ray beams with MeV energy resolution in a compact design. In its basic configuration, the spectrometer exploits conversion of gamma-rays into electrons via Compton scattering in a low-Z material. The scattered electron population is then spectrally resolved using a magnetic spectrometer. The detector is shown to be effective for gamma-ray energies between $3$ and $20$ MeV. The main properties of the spectrometer are confirmed by Monte-Carlo simulations.  
\end{abstract}

\maketitle

\section{\label{sec:IntroLidet}Introduction}

MeV photons are generated in a wide range of physical scenarios, ranging from unstable nuclei up to massive stellar objects such as pulsars. Effective detection and analysis of the information contained in the photon spectra have resulted in important discoveries in the fields of nuclear physics \cite{Nucleardiscovery} and astronomy \cite{Astrodiscovery}, and the development of new sources are important for many applications including security, nuclear physics, and medicine \cite{Nuclearcarwash,Nuclearcarwash2,comptom}. Accurate and efficient detection of high-energy photons is thus an important area of research for fundamental as well as for applied physics.

Currently there are a number of ways of detecting MeV gamma rays, most of which exploiting either Compton scattering or the quantum electromagnetic cascade initiated by the photon in a solid. Compton cameras are compact devices that detect gamma rays by measuring the number of electrons stripped off an atom in the crystal lattice. These free electrons impact onto a scintillator whose photoemission is then collected by a photomultiplier tube and counted. While Compton cameras can give accurate information about flux, they do not give information about the energy of incident photons \cite{ComptonCamera,ComptonCamera2}. On the other hand, the energy of a multi-MeV photon can be measured by detecting the quantum electrodynamic cascade induced during its passage through a material. The size and depth of the cascade yields a direct measurement of the energy of the incident photon \cite{cascadedet}. These devices can be extremely accurate but they suffer from the limitation that only single hit events can be precisely resolved.

Detectors have been developed which are able to successfully determine the energy of gamma rays as well as their flux simultaneously. These include the EUROBALL cluster \cite{euroball} which can spectrally resolve energies up to $10$ MeV, and more recently the AFRODITE germanium detector array, which can measure gamma rays with energies up to $20$ MeV \cite{afrodite}. Such detectors track the energy of the gamma by looking at its scattering through a volume of germanium. Information from the scattered electrons can be used to reconstruct the energy of the original gamma-ray. While such detectors are able to perform their tasks with great proficiency, the nature of their tracking means that a large number of germanium arrays are used in the device to build up a picture of flux and that increases the volume they occupy, making their implementation in many laboratories infeasible.	\\
\indent Here we report on a novel design that allows for simultaneous measurements of energy and flux of a multi-MeV gamma-ray beam in a compact detector design. The proposed detector exploits Compton scattering of the gamma-ray photons in a low-Z solid (Li in this case). It is shown that, for a suitable Li thickness, the scattered electron population retains a similar spectral shape as the initial gamma-ray beam, as long as the latter has an energy not exceeding $50$ MeV. By measuring flux and spectrum of this electron population by means of a magnetic spectrometer, it is thus possible to deconvolve the signal, retrieving the spectrum of the gamma-ray beam. The main properties of this detector are confirmed by proof of principle experiments \cite{LiF} and Monte Carlo simulations performed using the code FLUKA \cite{FLUKA1,FLUKA2}. \\
\indent The structure of the article is as follows. Section \ref{sec:GammaintLit} will briefly describe the main interaction mechanisms of a multi-MeV gamma-ray beam with a low-Z solid and introduce lithium as the best material for this detection system. Section \ref{sec:Liafterinteraction} will discuss the propagation of secondary particles (photons, electrons, and positrons) after they escape the lithium block and address the signal-to-noise ratio of the detector. This section will also describe the requirements of the magnetic spectrometer for the measurement of the spectrum of the scattered electrons. Finally, in Section \ref{sec:resultsLi} an overall design of the detector will be presented.

\section{\label{sec:GammaintLit}MeV Gamma-Ray Interaction with Lithium}

\begin{figure}[h!]
\includegraphics[scale=0.11]{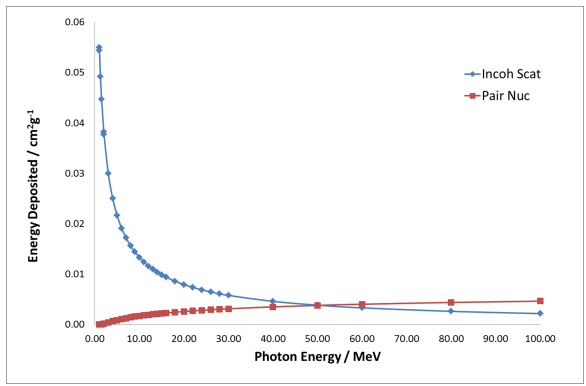}
  \caption{Energy loss for a multi-MeV gamma ray beam as it interacts with lithium. The red line with square points corresponds to pair production in the nuclear field, whereas the blue line with diamond points corresponds to Compton scattering. Compton scattering is dominant up to approximately $50$ MeV, beyond which pair production dominates. (Data obtained from the NIST database\cite{NIST}).}
  \label{fig:RespLi}
\end{figure}

\begin{figure}[h!]
\includegraphics[scale=0.64]{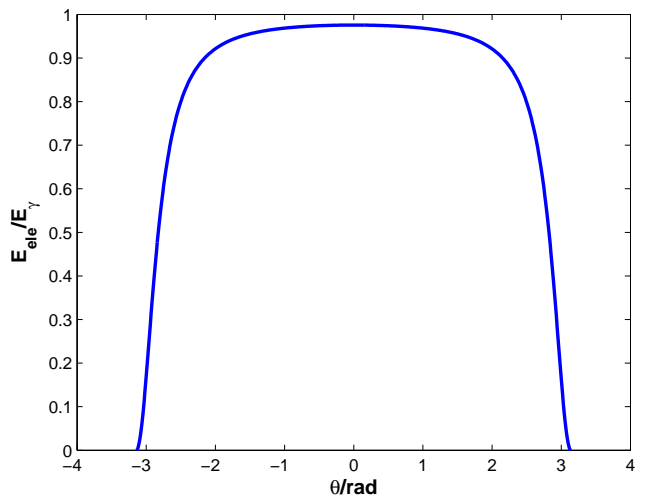}
  \caption{Angular dependence of the energy of Compton scattered electrons for  photon energy of $10$ MeV. For multi-MeV photons, the  electron energy remains largely constant over a large angular range, allowing the energy of the incoming $\gamma$-ray to be inferred from the Compton scattered electrons.}
  \label{fig:ScatangLi}
\end{figure}

Photons may interact with atoms in a number of ways. In the multi-MeV regime, the main interaction mechanisms are photoionization, Compton scattering, and electron-positron pair production. We will hereafter neglect ionization, since the energy loss associated with this mechanism is in the eV range i.e., much smaller than the initial photon energy and focus our attention only on Compton scattering and pair production. Compton scattering between an MeV photon and an electron at rest can be considered to be an essentially elastic process. The forward scattered electron is expected to have an energy comparable to that of the incident photon. On the other hand pair production is most likely to occur in the electromagnetic field of the nucleus. In this case, the process is intrinsically inelastic as momentum can be transferred in a way that allows a large spread of energies (3-body system). In order to generate an electron population with a spectrum similar to the incident gamma-ray beam, we thus need to maximize the probability of Compton scattering if compared to that of pair production. Since pair-production scales with the square of the atomic number Z \cite{LLClassicalFields1980,Sarripositron}, it is intuitive to expect that an element with low Z would be required. We will thus focus our attention on lithium, which is the material with lowest Z that is solid at standard conditions. Based on the NIST database \cite{NIST}, we can see how Compton scattering is indeed the dominant interaction mechanism for a multi-MeV gamma-ray beam up to energies of $50$ MeV (see Fig. \ref{fig:RespLi}).
\\
\indent Equation 1 shows the Compton scattering formula.
\begin{equation}\label{eq:Compton}
\Delta\lambda=\frac{h}{cm_e}(1-\cos\theta)
 \end{equation}
Where $\Delta\lambda$ is the change in photon wavelength, $h$ is Planck's constant, $c$ is the speed of light $m_e$ is rest mass of the electron and $\theta$ is the angle of the scattered photon. From equation \ref{eq:Compton} it is trivial to calculate the energy transferred to the electron. Fig. \ref{fig:ScatangLi} shows that for high energy photons, the energy of the Compton-scattered electrons is insensitive to the scattering angle over a wide range, forming the basis for the spectrometer design. 

While the primary Compton-scattered electrons are well correlated to the photon energy, good conversion efficiency into Compton electrons necessitates thick lithium converters. In order to determine the effect of the lithium converter on the the spectrum of the electrons escaping the lithium, a series of simulations were performed using the Monte-Carlo code FLUKA \cite{FLUKA1,FLUKA2}. For these simulations we assume pencil-like monoenergetic photon beams of different energy $E_{\gamma}$ ($1.5$, $2$, $5$, $7$, $10$, $15$, and $20$ MeV) incident upon $2$ cm of lithium. Scattering of the electron in the lithium broadens the electron spectrum and results in a low energy tail. This effect can be mitigated by selecting a narrow acceptance angle. This was investigated by simulating $10$ MeV photons incident on the $2$ cm lithium and varying the acceptance angle of the electrons detected immediately after it. 

\begin{figure}[h!]
	\includegraphics[scale=0.5]{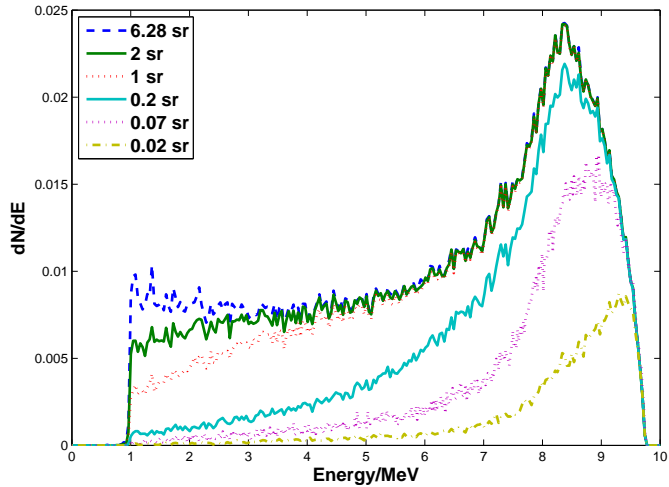}
  \caption{Spectra of the electrons generated after the propagation of a pencil-like monoenergetic gamma-ray beam of $10$MeV through $2$cm of lithium for various acceptance angles (see legend). The y-axis corresponds to the number of electrons generated per incident gamma-ray photon per MeV.}
  \label{fig:angleLi}
\end{figure}

From Fig. \ref{fig:angleLi}, it is clear that the shape of the peak is maintained for acceptance  angles between $2\pi$ and $1$ sr. With an acceptance angle of $0.2$ steradians, the level of the low energy tail significantly reduces with only a minor reduction in the peak conversion factor. The focus of attention in the remainder of this paper will be on electrons forward scattered in narrow cones that give a reasonable compromise between spectral resolution and high signal yield ($0.07$ steradians). 

\begin{figure}[h!]
	\includegraphics[scale=0.5]{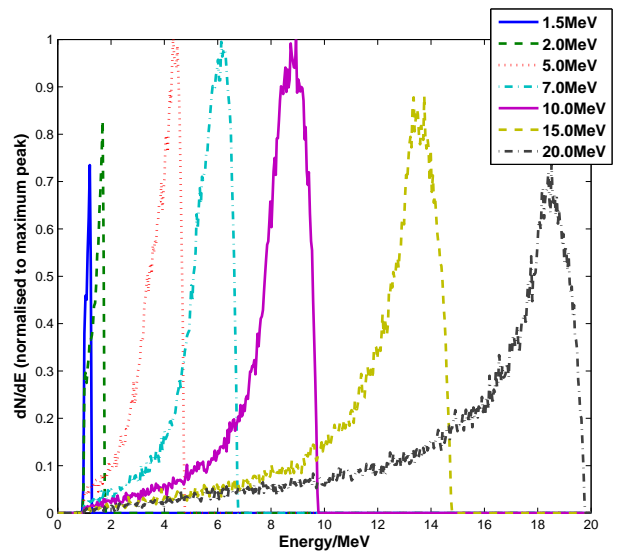}
  \caption{Spectra of the electrons generated after the propagation of a pencil-like monoenergetic gamma-ray beam through $2$cm of lithium (see legend for the different gamma energies simulated). The y-axis corresponds to the number of electrons generated per incident gamma-ray photon per MeV, normalized to the maximum peak. The angle of acceptance for electrons was $70$ msr.}
  \label{fig:relrespLi}
\end{figure}

Now we consider the dependence of the electron spectra on the incident photon energy. We find  a peaked distribution with a maximum at an energy approximately $1.6$ MeV less than the incident photon energy $E_{\gamma}$. The  full width half maximum increases with increasing photon energy and stabilises at a width of the order of $2$ MeV  at high energies plus a low energy linear tail for a $2$ cm thick Li converter. The simulated spectra of this electron population for a range of initial gamma-ray energies are shown in Fig. \ref{fig:relrespLi}.

The spectral broadening and the emergence of the low energy tail is largely due to the straggling of electrons as they move through the converter. Thus a trade-off  exists between spectral resolution and yield. 

\begin{figure}[h!]
	\includegraphics[scale=0.45]{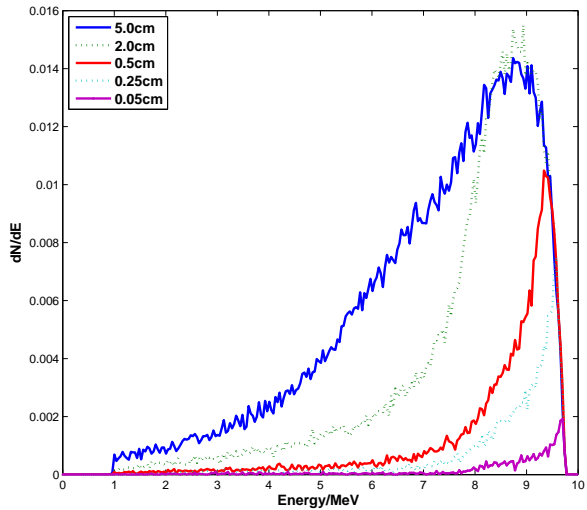}
  \caption{Spectra of the electrons generated after the propagation of a pencil-like monoenergetic gamma-ray beam of $10$MeV through a thickness of $5$ cm, $2$ cm, $0.5$ cm, $0.25$ cm and $0.05$ cm of lithium (see legend for the target thickness simulated). The y-axis corresponds to the number of electrons generated per incident gamma-ray photon per MeV. The angle of acceptance for electrons was $70$ msr. As the thickness of the lithium decreases, so too does the FWHM of the peaks. This comes at the cost of reduction to overall yield.}
  \label{fig:thicknessLi}
\end{figure}

Fig. \ref{fig:thicknessLi} shows the effect the various thickness of lithium for an interaction with $10$ MeV photons. The thinner converters achieve a significantly narrower electron peak ($<0.5$ MeV) at the cost of a noticeable reduction in yield. Note that at lower photon energies  the overlap of the response curves limits the achievable resolution and  that alternative systems can be used in this energy range. 

Using thin converters, deconvolution techniques allow for suitable energy resolution, while acceptable resolution at higher yields can be achieved with thicker converters. The thickness of lithium will largely depend on the flux of photons the user wishes to investigate. With high fluxes, a thinner target can be chosen; for our application  a compromise of $2$ cm was chosen.
\cite{LEDet}.
   
\begin{figure}[h!]
\includegraphics[scale=0.55]{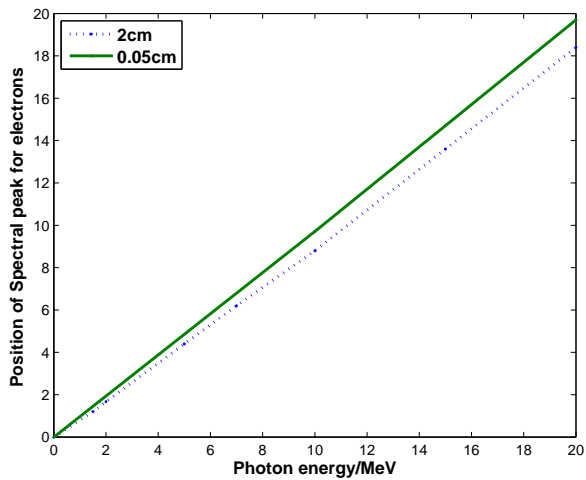}
  \caption{Fluka simulations show a linear relation between the initial photon energy $E_{\gamma}$ and the energy of the electrons escaping  the lithium converter. The position of the peak is affected slightly by scattering in the converter as shown for  two converter thickness of $2$ cm (blue dashed) and $0.05$ cm (green solid).}
  \label{fig:peakposLi}
\end{figure}

The simulations indicate a conversion efficiency into electrons of the order of $1.5$ to $3$\% for a $2$ cm target, depending on $E_{\gamma}$.
Fig. \ref{fig:peakposLi} shows the peak of the electron spectrum as a function of $E_\gamma$ on a $2$ cm thick target. 

\begin{figure}[h!]
	\includegraphics[scale=0.12]{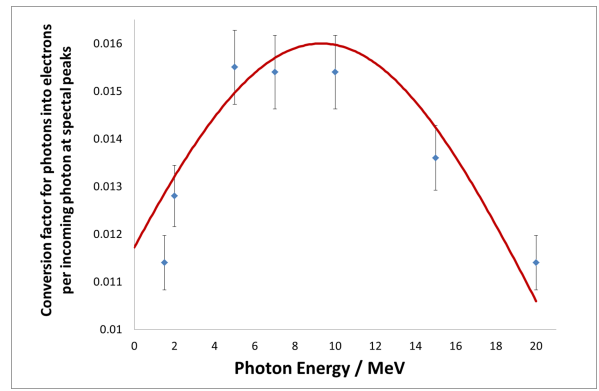}
  \caption{The peak of the conversion factors in the electron spectra plotted as a function of the initial photon energy. The peaks increase in size up to $5$ MeV where the peaks remain constant up to $10$ MeV after which they begin to decrease again. A Gaussian fitting is applied to the trend.}
  \label{fig:peaksizeLi}
\end{figure}

 The peak position for the $2$ cm thick target is well approximated in the region of interest by the function;                           
\begin{equation}\label{eq:Lipeakfit}
 E_p=aE_{\gamma}+b
\end{equation}
where $E_p$ is the energy of an electron at the peak of the spectrum and the constant $a$ and $b$ are, respectively: $a=0.93\pm0.01$, $b=(-0.25\pm0.01)$ MeV. The number of electrons produced per MeV per incident photon at peak energy is shown as a function of $E_{\gamma}$ in Fig. \ref{fig:peaksizeLi}.

The yield of electrons produced increases as the energy increases up to $5$ MeV where it begins to plateau before once again decreasing. This is likely due to the electrons having absorbed higher energy and hence being scattered less before the conversion efficiency drops. Within the region of validity (between $3$ and $20$ MeV) a Gaussian gives a suitable approximation of the relationship between the peak value of the conversion factor to within $5$\%:
\begin{equation}
\frac{dN_p}{dE}=C\exp[-\big(Κ(E_{\gamma}-D)\big)^2]
 \label{eq:Lifitpeak} 
\end{equation}                                   
where $N_p$ is the peak conversion factor value for electrons per incident photon in MeV$^{-1}$, $C=0.016\pm0.001\text{ MeV}^{-1}$ and $D=9.3\pm0.5$ MeV. For poly-chromatic gamma-rays the expected spectrum will be a superposition of electron spectra, it is thus possible to parametrize the expected electron spectrum in terms of the flux and energy of the incident gamma-ray beam using the two equations provided. It thus follows that a deconvolution of the measured electron spectrum is possible to allow one to retrieve the spectrum and flux of the incident photons.  It must be noted that, whilst performing the signal deconvolution, the low-energy tail of the electron spectrum must be taken into account. As a first approximation, a linear fit well parameterises this feature and it must be included for an accurate reconstruction of the gamma-ray spectrum.

\begin{figure}[h!]
	\includegraphics[scale=0.6]{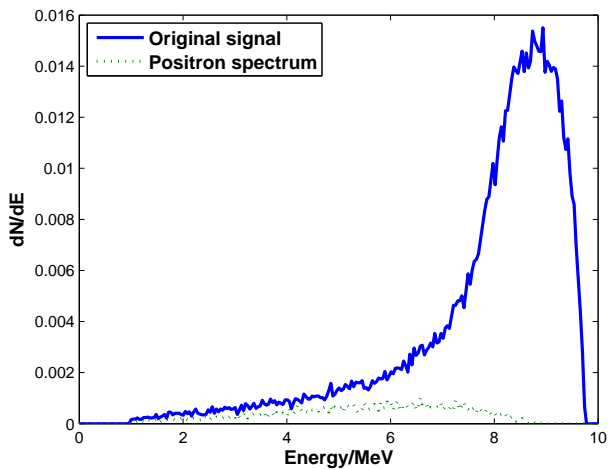}
  \caption{Spectrum of the electrons and positrons escaping from the rear surface of the lithium block of $2$cm thickness once a pencil-like monoenergetic gamma-ray beam with an energy of $10$ MeV is incident on it. Results are expressed in number of particles generated per incident gamma-ray photon per MeV.}
  \label{fig:eleposLi}
\end{figure}

As a final remark, it is worth noting that, even if probability of pair production is significantly smaller than that of Compton scattering for $1<E_{\gamma}$[MeV]$<20$ (see Fig. \ref{fig:RespLi}), neglecting pair production effects  introduces a small systematic error in extracting the photon spectrum once the spectrum of the scattered electrons is known. As an example, we plot in Fig. \ref{fig:eleposLi} the spectra of the electrons and positrons escaping from the rear surface of the lithium once a pencil-like monoenergetic gamma-ray beam with an energy of 10 MeV is incident upon it.

As we can see from Fig. \ref{fig:eleposLi}, a small population of positrons is indeed generated. Integrating the positron and electron spectrum, we can see that the approximately  $12$ times more electrons than positrons escape the $2$ cm thick lithium target. For thinner converter targets ($0.5$ cm and $0.05$ cm), the low energy tail is almost entirely due to pair production and it is therefore possible to remove it entirely by using a thin lithium converter as long as the flux of incident gammas remains sufficiently high. 

FLUKA simulations indicate the spectra of electrons and positrons arising from pair production to be very similar, as theoretically expected \cite{LLClassicalFields1980}. The energy of the positrons is slightly higher than that of the electrons created via pair production (Fig. \ref{fig:eleposLi} shows a positron peak at $6$ MeV from a monoenergetic gamma-ray beam of $10$ MeV). It is worth noticing that, whilst very similar, the spectra of the electrons and positrons arising from pair production are not exactly the same. A complete symmetry in pair production holds only in the ultra-relativistic regime (Born approximation), i.e. when both the electron and positron energies greatly exceed the electron's rest mass. This is not strictly true in the mildly relativistic regime considered here, whereby the 3-body Coulomb interaction between electron, positron, and nucleus should be taken into account. However, these corrections are small(of the order of a few percent) and can be neglected here. It is thus a reasonable approximation to subtract the positron spectrum from the electron one, in order to extract the spectrum of electrons uniquely arising from Compton scattering \cite{LLClassicalFields1980}.\\
\indent A simultaneous measurement of the spectra of the electron and positron populations (see Section \ref{sec:resultsLi}), allows for  subtraction of the latter from the electron spectrum, thus largely eliminating this source of systematic error (details in section \ref{sec:Liafterinteraction}).

\section{\label{sec:Liafterinteraction}Behaviour of Particles After Interaction with Lithium}

\begin{figure}[h!]
	\includegraphics[scale=0.7]{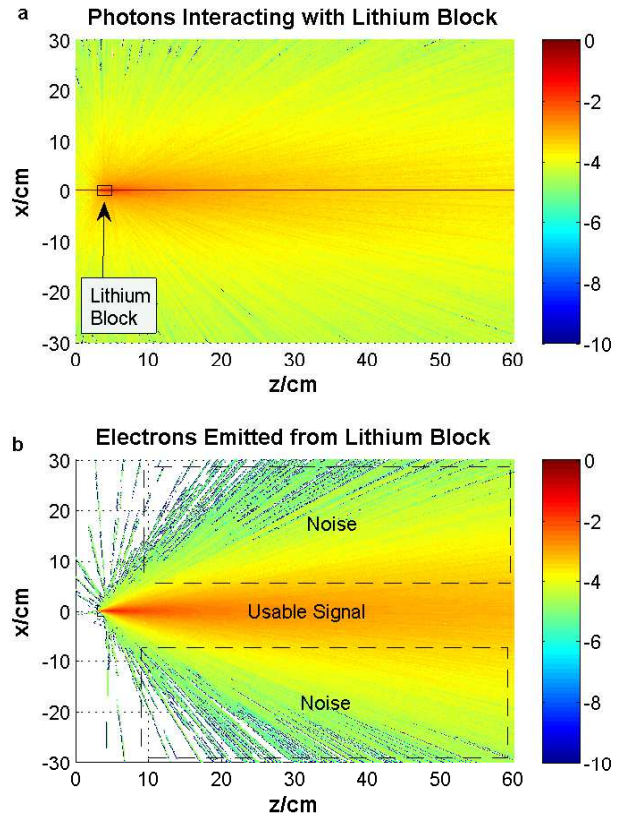}
  \caption{Spatial distribution of photons (a) and electrons (b) after the interaction of a $10$ MeV gamma-ray beam with a $2$ cm thick lithium cube. In both cases, the colourbar represents the number of particles per incident photon per cm (x-axis) expressed as a base $10$ logarithm.}
  \label{fig:noiseLi}
\end{figure}

In the previous section, we have focused our attention exclusively on the electrons which are emitted on the same axis as the incident gamma-ray beam. This is in principle over-simplifying, since it is well known that Compton scattering present a broad angular distribution (Fig. \ref{fig:ScatangLi}). Electrons will thus be emitted also at wider angles, with an energy that gradually decreases as we move far from the axis of the gamma-ray beam. Those electrons thus represent a source of noise for the detector and should be suppressed. Compton scattering will also induce a broadly divergent population of scattered photons, which are again a detrimental source of noise when attempting to measure the spectrum of the scattered electrons. These qualitative arguments are quantitatively corroborated by Fig. \ref{fig:noiseLi}, which show the spatial photon (\ref{fig:noiseLi}a) and electron (\ref{fig:noiseLi}b) distributions, as resulting from FLUKA simulations, once a pencil-like monoenergetic gamma-ray beam with an energy of $10$ MeV interacts with a $2$ cm thick lithium block. 

\begin{figure}[h!]
	\includegraphics[scale=0.7]{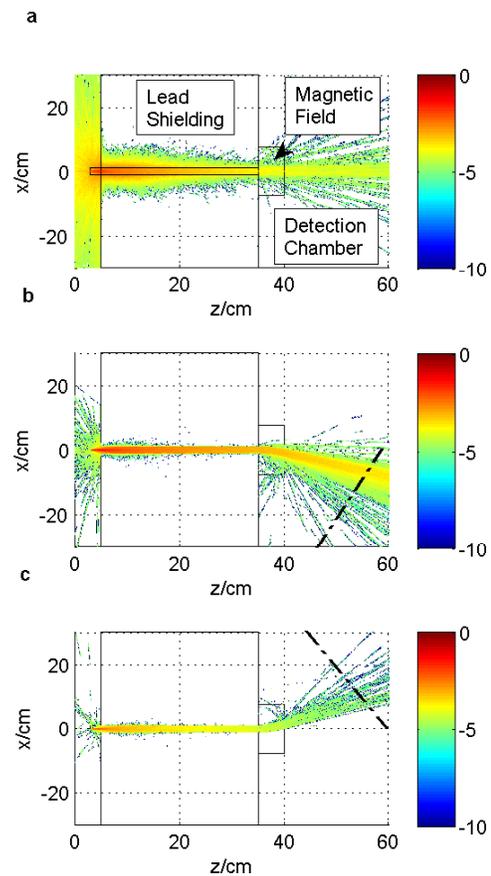}
  \caption{Spatial distribution of photons (a), electrons (b) and positrons (c) after the interaction of a $10$ MeV gamma-ray beam with a $2$ cm thick lithium cube once the lead shielding and a $5$ cm-long $0.3$ T magnetic field is inserted. The colourbar represents the number of particles per incident photon per cm (x-axis) expressed as a base $10$ logarithm. The dashed lines in b and c indicate the possible locations of an image plate or LANEX screen.}
  \label{fig:detectorresponseLi}
\end{figure}

As we can see the vast majority of photons pass through the lithium block without interacting (less than 95\% of the initial photons). Also, a population of scattered photons will be isotropically emitted by the lithium (approximately $10^{-3}$ or $10^{-4}$ photons per incoming gamma-ray photon). The situation is analogous for the scattered electrons. We can see from Fig. \ref{fig:noiseLi}b that the electron spatial distribution consists of two main parts: a relatively strong population of forward scattered electrons (labelled as usable signal in Fig. \ref{fig:noiseLi}b) and a wider distribution of electrons scattered widely at both sides of the lithium (labelled as noise in Fig. \ref{fig:noiseLi}b). This latter population must be suppressed if a good signal-to-noise ratio is to be achieved with the detector. The easiest way to deal with this noise is to introduce a thick lead shielding immediately adjacent to and at either side of the lithium. If we assume a $30$ cm thick lead wall with an aperture in the middle of the same size as the lithium block, almost all the off-axis scattered electrons and photons are effectively absorbed in the lead. This is shown in Fig. \ref{fig:detectorresponseLi} which depicts the simulated photon, electron, and positron spatial distribution for the same initial conditions as the ones used for Fig. \ref{fig:noiseLi}, with the only difference that a $30$ cm lead wall has been now inserted at each side of the lithium block.

Whilst, the photon signal on axis has obviously not been affected by this shielding, the off-axis noise has dropped by several orders of magnitude (see Fig. \ref{fig:detectorresponseLi}). If we now insert a magnetic field at the rear of the lead shielding, we can effectively separate, and spectrally resolve, the electrons and the positrons (see Fig. \ref{fig:detectorresponseLi}). The spectrally separated electrons and positrons can then be recorded on a suitable, spatially resolved detector, such as an Image Plate or a LANEX screen placed at locations indicated by the dashed lines in figure \ref{fig:detectorresponseLi} b and c \cite{IPcal1,IPcal2,LANEX}. The spectral resolution of this magnetic spectrometer will depend upon a series of parameters, namely the strength $B$ and spatial extent $L_m$ of the magnetic field, the angular divergence $\theta_s$ and source size $D_s$ of the beam, and the distance $D_d$ between the entrance of the magnet and the detector. The relative energy resolution $\delta E/E$ of a magnetic spectrometer in the ultra-relativistic regime $(E\gg0.5 \text{MeV})$ thus reads \cite{PhDmag}
\begin{equation}
\frac{\delta E}{E}\approx \frac{E(eV)}{cB(T)}\frac{(D_s+D_d)\theta_s}{(D_d+L_m/2)L_m}
 \label{eq:Liresolution}
 \end{equation}
Ideally a small aperture, a strong magnetic field, and a long distance between the source and the detector are advisable in order to improve the spectral resolution. However, it is clear that all these parameters significantly decrease the yield of particles on the detector. A trade-off must then be found between a good spectral resolution and a good signal on the detector. Calculations show that a reasonable compromise is found for $B=0.3\text{ T}$, $L_m=5\text{ cm}$, $\theta_s\approx30\text{ mrad}$, $D_s=1\text{ cm}$, $D_d=25\text{ cm}$ (see Fig. \ref{fig:noiseLi}). 

\begin{figure}[h!]
	\includegraphics[scale=0.5]{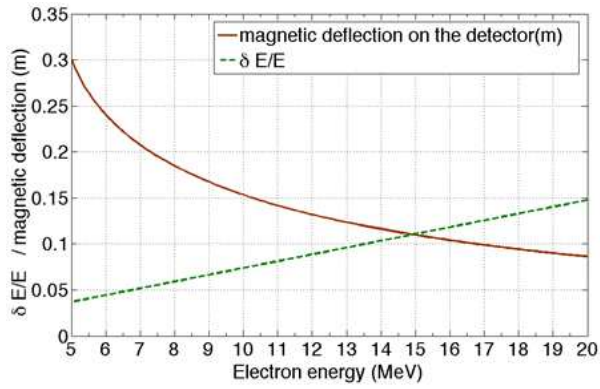}
  \caption{Relative energy resolution (dashed green line) and spatial deflection on the detector (solid brown line) of the scattered electron population after propagation through a $5$cm long, $0.3$T magnet. Details in the text. }
  \label{fig:resolutionLi}
\end{figure}

As expected, the energy resolution decreases as the electron energy is increased. However, a relative energy resolution of $4$\% and $12$\% are found for $5$ MeV and $15$ MeV, respectively (energy resolution of $0.2$ MeV and $1.8$ MeV, respectively). This energy resolution is less than or comparable to the intrinsic energy resolution during the conversion of gamma-ray photons to electrons in the lithium target (see Fig. \ref{fig:resolutionLi}). The resolution of the magnetic spectrometer can also be improved by positioning the detector further away, but at a cost of reduced compactness.
As a final remark, it is interesting to note that the vast majority of the gamma-ray photons are able to escape unperturbed from the detector (more than $95$\%). This allows for simultaneous detection of the spatial profile of the gamma-ray beam, if a suitable detector is placed on axis.
\section{\label{sec:resultsLi}Final Design and Conclusions}
\begin{figure}[h!]
	\includegraphics[scale=0.12]{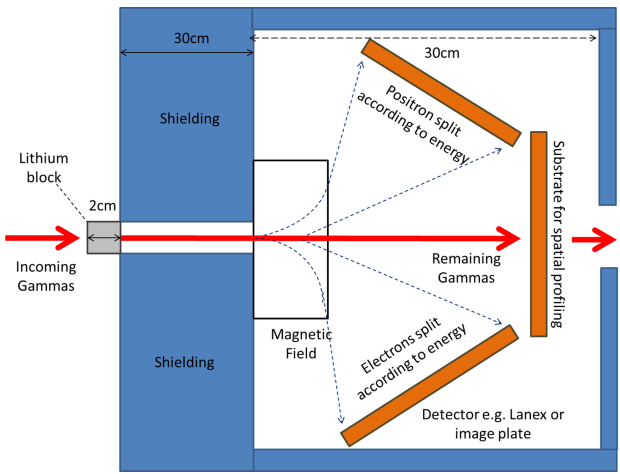}
  \caption{The proposed setup for the gamma-ray spectrometer (not to scale). The entire device is $60$ cm long in total and consists of $30$ cm of lead shielding with a $1$ cm thick gap at the centre before which the Lithium sits. A magnetic field of strength $0.3$ T is immediately adjacent to this. Image plates can sit off axis in order to detect the charges deflected. On axis, a substrate can be placed for spatial profiling.}
  \label{fig:finalmodelLi}
\end{figure}

A proposed setup for the detector is depicted in Fig. \ref{fig:finalmodelLi}. The gamma-ray photon to be measured is incident upon a $2$ cm thick lithium block surrounded by $30$ cm of lead shielding with a $1\times1$ cm$^2$ aperture. The entire device shown in Fig. \ref{fig:finalmodelLi}, including the shielding is $60$ cm long and hence, able to fit conveniently into most laboratories. A population of electrons and positrons is generated, the latter arising from pair production in the nuclear field whereas the first generated partially by pair production and mostly by Compton scattering. For a gamma-ray energy window of $1<E\text{[MeV]}<20$, Compton scattering is dominant (see Figs. \ref{fig:RespLi} and \ref{fig:eleposLi}) allowing for effects due to pair production to be neglected (less than $10$\% of the overall electron yield). 

The scattered electrons will have an energy distribution that resembles that of the incident gamma-ray beam (bell-shaped distribution plus a linear low-energy tail, see Fig. \ref{fig:thicknessLi}). The electron distribution can be parametrized as a function of the flux and energy distribution of the incident gamma-ray beam (see Figs. \ref{fig:thicknessLi} and \ref{fig:peakposLi}), allowing for deconvolution of the electron signal. Once the electron spectrum is measured by a magnetic spectrometer, the spectrum of the incident gamma-ray beam can thus be retrieved with a resolution of the order of the MeV (see Fig. \ref{fig:resolutionLi}). The lead shielding allows for fine selection of the scattered electrons and reduces the off-axis noise induced by photons and particles scattered at a wide angle (see Figs. \ref{fig:noiseLi} and \ref{fig:detectorresponseLi}). 

In conclusion, the design for a compact gamma-ray spectrometer suited for high-flux gamma-ray beams is presented. The system has an energy resolution of the order of one MeV in an energy window of $3-20$ MeV. The performance of the detector is analysed by Monte-Carlo simulations. The system has been tested in  recent experimental campaigns (details of the experiment to be published elsewhere \cite{Sarri2Beam}) confirming the numerical predictions presented in this manuscript.  

\begin{acknowledgements}
The authors are grateful to M. Montgomery for his technical support throughout the project. D.J. Corvan wishes to acknowledge financial support from EPSRC (grant number EP/L013975/1). G. Sarri wishes to acknowledge the financial support from the Leverhlume Trust (grant number ECF-2011-383).
\end{acknowledgements}

\bibliographystyle{unsrt}

%\bibliography{Bibliography}

\end{document}